\documentclass[prl,twocolumn,showpacs]{revtex4}  
\usepackage{graphicx}  

\begin{document}


\title{Resonance Scattering of Fast Electrons in a Single Crystal
\footnote{ JETP Letters, 1979, Vol. 29, No.6, pp. 302-307} }

\author{G.\,V.\, Kovalev,  N.\,P.\, Kalashnikov \/\thanks}



\affiliation{Engineering-Physics Institute, Moscow }


\date{Jan. 12, 1979}

\begin{abstract}
It is shown that resonance scattering of fast ($pL > >pR > > 1$, $p$ is the particle 
momentum, $L$ is the longitudinal dimension of the potential, and $R$ is its 
transverse dimension) charged particles occurs in the extended potential. The 
data on small-angle scattering of fast electrons in the crystal are interpreted on 
the basis of the examined effect.
\end{abstract}

\maketitle


It is generally assumed that resonance scattering in elastic collisions applies to 
slow particles $pR <<1$[1]. As will be shown below, however, resonances in the elastic 
scattering can also be present in fast particles, i.e., when $pR>1$, if the particle is
scattered by an extended, attracting potential. Physically, this effect consists of the 
following, the cross section of the extended, attracting potential has bound and quasi-bound states and hence the transverse component of the incident wave of a fast particle may undergo resonance scattering on these states. In this case “the slow” transverse motion of the particle is a necessary requirement, i.e., $pR<<1$. This imposes a 
constraint on the entrance angle of the particle $\theta_0<<1/pR$.

Let us examine the scattering by a string potential which has the appearance of a 
square well (Fig. 1):
\begin{eqnarray}
 U( \vec{\rho}, z)= \left \{\begin{array}{ll} -V_o, & \text{for} \; \rho  \leq R , 0 \leq z \leq L, \\  
 0,&   \text{for}  \; \rho > R , z < 0,  z > L,
\end{array} \right. 
\label{r1}
\end{eqnarray}
where $L$ is the length of the string and $2R$ is its transverse diameter ($L>>2R$ ). If we 
assume that a fast particle  $p L> p R > 1$ enters almost parallel to the $z$ axis at an angle $theta_o$,  then from the condition for joining the wave function of a particle at the boundary  $z = 0$ and $z = L$ and for the asymptotic behavior at infinity we obtain the scattering amplitude [2]
\begin{eqnarray}
f(\theta, \phi)=\\ \nonumber
\frac{p}{2 \pi i} \sum Q_{p_{\perp i} }(\alpha, m) Q^{*}_{p_{\perp f} }(\alpha, m) [ \exp(i \frac{\alpha-p_{\perp }^{2}}{2 p}L) -1].
\label{r2}
\end{eqnarray}
where
\begin{eqnarray}
Q_{p_{\perp } }(\alpha, m) =\int d^2 \, \vec{\rho}\, Z_{\alpha, m} ( \vec{\rho} ) \exp(i p_{\perp } \vec{\rho} ).\nonumber 
\label{r3}
\end{eqnarray}
is the amplitude of the transition from the state of the plane wave with the momentum 
$p_{\perp } $  to the eigenstate $Z_{\alpha,m}(\vec{\rho})$ of the transverse motion in the potential (1); the summation is carried out over the compound states with the transverse energy $\epsilon(n,m) = \alpha /2p$ and momentum $m$. To calculate the total cross section we use the optical theorem  $\sigma =(4 \pi /p) Im[f(\theta, \phi)]$.
We obtain
\begin{eqnarray}
\sigma =4 \sum | Q_{p_{\perp i} }(\alpha, m) |^2 \sin^2 (\frac{\alpha-p_{\perp }^{2}}{2 p}L) .
\label{r4}
\end{eqnarray}
It was shown earlier[3] that, as a result of scattering of a fast particle by a sufficiently 
extended potential ($L>>pR^2$), the effective scattering angle becomes very small 
$\theta_{eff} \approx  1 / \sqrt{L/p}$ and the corresponding effective impact parameter very large $\rho_{eff} \approx  1 /p \theta_{eff} \approx \sqrt{L/p}$.

\begin{figure}
	\centering
		\includegraphics[width=0.450\textwidth]{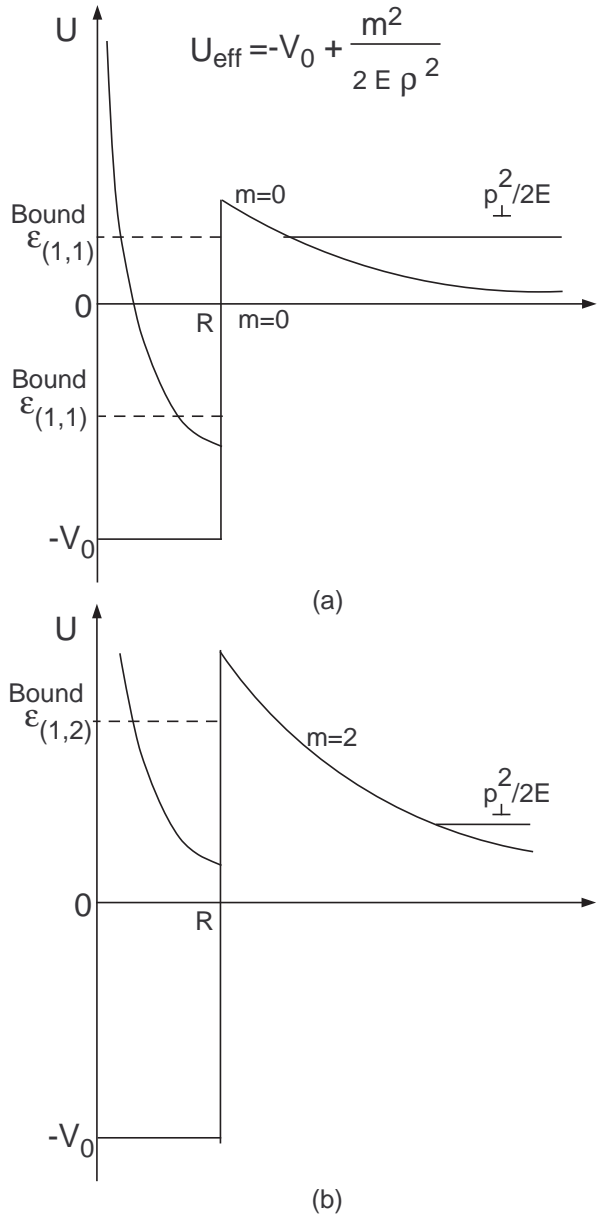}
	\caption{Transverse cross section of the string’s potential.}
	\label{fig:fig_3}
\end{figure}

The particle with the azimuthal momentum $m$ passes the string at a 
distance of $\approx m/p_{\perp }$ and hence is scattered by the potential (1) if this distance is smaller than the effective impact parameter
\begin{eqnarray}
\frac{m}{p_{\perp i}} \leq \rho_{eff} =\sqrt{\frac{L}{p}} .
\label{r5}
\end{eqnarray}
If
\begin{eqnarray}
\frac{1}{p_{\perp i}}> \sqrt{\frac{L}{p}} \,\,\,  (\theta_0< \frac{1}{\sqrt{pL}}  ) , \nonumber
\label{r6}
\end{eqnarray}
then only the wave with $m = 0$ is scattered. In this case the scattering cross section (3), 
which is integrated and summed over the intermediate states, has the following form:
\begin{eqnarray}
\sigma=\sigma_0+ \frac{2}{\pi} \sum \frac{|\epsilon^{B}(n,0)|}{p} \frac{\sin^2[(|\epsilon^{B}(n,0)|+\frac{p^{2}_{\perp i}}{2p})\frac{L}{2}]}{[|\epsilon^{B}(n,0)|+\frac{p^{2}_{\perp i}}{2p}]^2} .
\label{r7}
\end{eqnarray}
where $ \sigma_0 \approx (2/\pi) L /p$ is the scattering cross section from the continuous spectrum with $
m = 0$, which was calculated elsewhere, [3] and the second term is determined by the 
bound states $[\epsilon^{B}(n,0)<0]$ in the transverse potential of the attracting string. We can easily see that the main contribution to the scattering cross section is introduced by 
bound states which are removed a distance of $\approx 1/L$ from the edge of the well. At the same time, the scattering cross section has a narrow peak $\approx 1/\sqrt{pL}$ in width near the zero angle of entry.

Let us assume now that the angle of entry is $1/\sqrt{p L} < \theta_0 < 1/p R$. The effective number of waves participating in the scattering in this case is determined by the inequality (4). It follows from relation (5) that the bound states $[ \epsilon^{B}(n,m) < 0]$  can be disregarded. However, for partial waves with $m > 0$ because of the presence of a centrifugal barrier $m^2/2 E R^2$ quasi-bound states with $\epsilon^{B}(n,m)$ are developed in the effective potential (Fig. 1). Assuming that the barrier is infinitely high and substituting the  square well for the effective well, we obtain
\begin{eqnarray}
\epsilon^{B}(n,m)=\frac{\pi^2 n^2}{2 E R^2}+\frac{m^2}{2E R^2} - V_0.
\label{r8}
\end{eqnarray}
Taking into account the finite depth of the effective well  $(\approx V_0)$, we can calculate the number of quasi-bound states corresponding to the azimuthal momentum $m$
\begin{eqnarray}
n^2_{max}=\frac{2EV_0 R^2}{\pi^2}.
\label{r9}
\end{eqnarray}
It can be seen that $n_{max}$, which is independent of $m$, is determined by the parameters of the well and by the total energy of the particle. In the energy region $E < \pi^2/2 V_0 R^2$ (for silicon $E <  20$ MeV) there is only one quasi-bound state $\epsilon^{B}(1,m)$  for a given $m$. It follows from Eq. (6) that $\epsilon^{B}(1,m) >0$ . 
Moreover, we always have $\epsilon(n,m)> \epsilon(n,m-1)$  (Fig. 1b). If the transverse energy now coincides with the energy of a certain quasi­bound state (Fig. 1a)
\begin{eqnarray}
\frac{p^2_{\perp i}}{2 E}=\frac{\pi^2 n^2+m^2}{2 E R^2} - V_0,
\label{r10}
\end{eqnarray}
then we have resonance scattering. The scattering cross section, which can be determined from expression (3) by integrating over the continuous spectrum from $0$ to $ \infty $, in this case has the Breit-Wigner form
\begin{eqnarray}
\sigma=\sigma_0+\frac{2}{\pi} \frac{L}{p} \frac{\frac{1}{4}  \Gamma^2}{(\theta_0-\theta_{0 \, res})^2+\frac{1}{4}  \Gamma^2},
\label{r11}
\end{eqnarray}
where $\sigma_0$ is the cross section for scattering at some distance from the resonance. We can easily obtain $\theta_{0 \, res}$ from relation (8)
\begin{eqnarray}
\theta_{0 \, res}=\left(\frac{\pi^2 n^2+m^2}{ E^2 R^2} -2 \frac{ V_0}{E}\right)^{1/2}.
\label{r12}
\end{eqnarray}
Determination of the finite centrifugal barrier using the standard quasiclassical treatment [1] yields the following angular width of the resonance
\begin{eqnarray}
 \Gamma \approx \theta_{0 \, res} \cdot  \exp- |2E V_0 R^2 -\pi^2 n^2|^{1/2}.
\label{r13}
\end{eqnarray}
If the angle of entry remains constant and the energy of the particle is varied, then 
from Eqs. (8), (9), and (11) we obtain
\begin{eqnarray}
\sigma=\sigma_0+\frac{2}{\pi} \frac{L}{p} \frac{\frac{1}{4}  \bar{\Gamma}^2}{(E-E_{ res})^2+\frac{1}{4}  \bar{\Gamma}^2};  \nonumber  \\
\bar{\Gamma} \approx E_{ res} \cdot  \exp - |2 E_{res}  V_0 R^2 -\pi^2 n^2|^{1/2}; \nonumber \\ 
E_{res} = \theta_{0}^{-2} \left[2 V_0 +  ( 4 V_0^2+ 4 \theta_{0}^{2}  R^{-2} (\pi^2 n^2 + m^2) )^{1/2} \right].
\label{r14}
\end{eqnarray}

The indicated resonance can be observed in the crystal when the fast particle 
enters it at a small angle to the crystallographic axis.

\begin{figure}
\centering
		\includegraphics[width=0.45\textwidth]{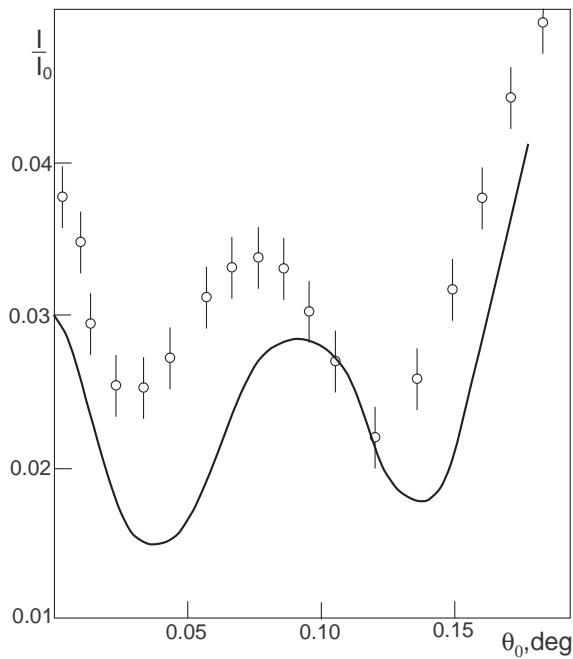}
\caption{The intensity of the 15-MeV electrons scattered in the $1.4-\mu$-thick silicon crystal as a function of  the angle of entry $\theta_{0}$ relative to the direction of the  $\left\langle 111 \right\rangle $ axis, $I_0$ is the intensity of the incident particles.  The experimental points were taken from Ref. 4; the solid curve denotes theoretical calculation.}
	\label{fig:fig_2}
\end{figure}

In this case the longitudinal momentum transfer becomes very small 
$ q_{\|} \approx 1/pR^2$ and the wave function of the particle, which is insensitive to the details of the behavior of the potential at approximately  the interatomic distances, is determined by a certain potential averaged over the length 
of the chain[2].   Schiebel and Worm[4] observed the scattering of 15-MeV electrons 
along the $\left\langle 111 \right\rangle $ axis of a $1.4-\mu$m-thick $Si$ crystal. Figure 2 shows a small-angle dependence of the intensity of the scattered particles on the angle of entry relative to the 
crystallographic $\left\langle 111 \right\rangle $ axis. This dependence cannot be explained in terms of the usual 
two-wave diffraction theory and the classical string scattering [5].  From the viewpoint 
of the discussion conducted above, the very narrow peak near the zero angle of entry is due to the true bound states; at the same time, its width $\approx 1/\sqrt{p L} \approx 0.02$degree is very close to experimental result.
Wide side peak is determined by scattering on the quasi-bound state with $n = 1$ and  $m = 1$. From (10) and (11) for the parameters of $Si $ $R^{-1} = m \, e^2 Z^{1/3} = 9.7 \cdot 10^3$ eV, $V_0 = 23$ eV, we obtain  $\theta_{0 \,res}\approx  0.1$ degree  and  $\Gamma \approx  0.5 \theta_{0 \,res}$  that is even more in line with the experimental results than would be expected from the evaluative expressions obtained.


\end{document}